\documentclass[conference]{IEEEtran}
\usepackage[utf8]{inputenc} 
\usepackage[dvips]{graphicx}
\usepackage{algorithmic}
\usepackage{algorithm}
\usepackage{array}
\usepackage{verbatim}
\usepackage{graphicx}
\usepackage{subfigure}
\usepackage[T1]{fontenc}
\usepackage{url}
\usepackage{ifthen}
\usepackage{cite}
\usepackage[cmex10]{amsmath} 
\usepackage{amssymb,amsthm,amsfonts}
\usepackage{color,changepage}
\usepackage{booktabs}

\newtheorem{Theorem}{Theorem}

\theoremstyle{remark}
\newtheorem{Remark}{Remark}

\newcommand{\tightpmatrix}[1]{%
  \begingroup
  \setlength{\arraycolsep}{3pt}
  \renewcommand{\arraystretch}{1}
  \begin{pmatrix}
    #1
  \end{pmatrix}%
  \endgroup
}

\newcommand{\tightbmatrix}[1]{%
  \begingroup
  \setlength{\arraycolsep}{3pt}
  \renewcommand{\arraystretch}{0.9}
  \begin{bmatrix}
    #1
  \end{bmatrix}%
  \endgroup
}

\begin{document}

\title{Secure Aggregation with Top-$K$ Sparsification in Decentralized Federated Learning} 

\author{\IEEEauthorblockN{Hengxuan Tang}
\IEEEauthorblockA{\textit{Southwest Jiaotong University} \\
hxuantang@my.swjtu.edu.cn}
\and
\IEEEauthorblockN{Jinbao Zhu}
\IEEEauthorblockA{\textit{Southwest Jiaotong University} \\
jinbaozhu@swjtu.edu.cn}
\and
\IEEEauthorblockN{Xiaohu Tang}
\IEEEauthorblockA{\textit{Southwest Jiaotong University} \\
xhutang@swjtu.edu.cn}
}

\maketitle

\begin{abstract}
Secure aggregation is a vital component for mitigating gradient leakage in federated learning, but its communication cost conventionally scales with the gradient dimension. This becomes prohibitive for large models and even more pronounced in decentralized federated learning with limited bandwidth and unreliable nodes. Top-$K$ gradient sparsification is an effective approach to reduce communication by transmitting only a few entries of the full gradient, while maintaining competitive model accuracy. Nevertheless, the top-$K$ entries selected by each user are unpredictable and vary across users, which poses a challenge for efficient sparse secure aggregation. This paper studies information-theoretic secure aggregation with top-$K$ sparsification in decentralized federated learning under user dropouts and user collusion. We propose a communication-efficient sparse secure aggregation scheme that offloads dimension-dependent overhead to an offline phase and protects private gradients using random masks and permutations. Experimental results demonstrate that our scheme preserves accuracy comparable to full-gradient aggregation even with only $1\%$ gradient sparsification, while substantially reducing the communication cost.
\end{abstract}

\section{Introduction}
Federated learning (FL) \cite{mcmahan2017communication} serves as a promising paradigm for distributed machine learning, enabling users to collaboratively train a global model by exchanging local updates (e.g., gradients) coordinated by a central server, while keeping their private data on-device.
Despite its privacy-preserving design, FL remains susceptible to gradient reconstruction attacks \cite{zhu2019deep, zhao2020idlg} that can reconstruct the original raw data from the gradients to some extent.
Secure aggregation \cite{bonawitz2017practical, so2021turbo, jahani2023swiftagg+} mitigates the risk of privacy leakage by only allowing the communication of encrypted gradients, without compromising model accuracy. In practical deployments, a major challenge for secure aggregation is user dropouts. Existing secure aggregation protocols address this issue, typically by introducing an additional communication phase to ensure that the aggregation can still be completed despite the dropout users \cite{zhao2022information,so2022lightsecagg,11176872}.

With the growing prevalence of edge deployments, decentralized federated learning (DFL) \cite{lalitha2018fully, beltran2023decentralized, yuan2024decentralized} has emerged as a serverless alternative to traditional FL. However, since DFL still relies on exchanging gradients among users, it inherits the privacy vulnerabilities of conventional FL, motivating the development of decentralized secure aggregation (DSA).
In particular, in recent years, the fundamental communication performance of DSA has attracted significant attention from the information-theoretic community.
The problem of information-theoretic DSA with a prescribed number of colluding users was initially introduced by Zhang \textit{et al.}~\cite{zhang2025information}, who characterized the optimal communication cost under the assumption that a trusted third party exists to share random keys among users. Subsequently, Li \textit{et al.}~\cite{li2025capacity} removed the reliance on a trusted third party and studied the DSA problem with a prescribed number of colluding users by allowing a group of users to share uncoded random keys.
More recently, Li \textit{et al.}~\cite{li2025optimal} further reduced the required key size by considering more general collusion patterns. 
Despite not considering user dropouts, these works \cite{zhang2025information, li2025capacity, li2025optimal} incur communication costs that scale with the gradient dimension, making them increasingly prohibitive for modern large-scale models.

Gradient sparsification is widely employed to alleviate the communication burden by restricting each user to transmit only $K$ entries of its local gradient \cite{aji2017sparse, lin2018dgc}.
Common sparsification strategies include random-$K$ sampling, which selects $K$ entries uniformly at random, and top-$K$ selection, which picks the $K$ largest-magnitude entries \cite{xu2021grace}. The information-theoretic secure aggregation with random-$K$ sparsification has been innovatively addressed by Sami and G{\"u}ler in \cite{sami2024secure}.
While top-$K$ sparsification achieves higher model training accuracy compared with random-$K$, information-theoretic secure aggregation with top-$K$ sparsification remains unexplored.
In contrast to random-$K$, where the indices of the selected $K$ entries are independent of the gradients, those selected by top-$K$ are inherently determined by the gradients themselves.
This dependence makes the design of privacy-preserving secure aggregation schemes with top-$K$ sparsification more challenging.

In this paper, we consider the problem of information-theoretic secure aggregation with top-$K$ sparsification in the DFL system. We propose a two-phase communication-efficient secure aggregation scheme that enables the aggregation of users’ top-$K$ sparse gradients, while simultaneously tolerating user dropouts and guaranteeing information-theoretic security against a prescribed number of colluding users.
Compared with traditional two-phase secure aggregation schemes \cite{so2022lightsecagg, zhao2022information, 11176872}, our approach reduces the first-phase communication to only $K$ masked entries, instead of full-dimensional masked gradients, 
and preserves the same second-phase communication as these existing schemes.
Compared with random-$K$ secure aggregation \cite{sami2024secure}, experimental results show that the proposed top-$K$ scheme attains superior model accuracy.

\subsubsection*{Notation}
For integers $m \leq n$, we write $[n] \triangleq \{1,2,\dots,n\}$ and $[m:n] \triangleq \{m,m+1,\dots,n\}$.
Let bold capital letters and bold lowercase letters represent matrices and vectors, respectively.
For a finite set $\mathcal{K}$, let $\vert \mathcal{K} \vert$ denote its cardinality. 

\section{Problem Formulation}
\label{sec:sysmodel}
Consider a decentralized federated learning (DFL) system consisting of $N$ users, where each user $n\in[N]$ holds a private input $W_n$ (e.g., its local gradient) of length $L$ over a finite field $\mathbb{F}_q$ of size $q$. The users are connected by a fully connected peer-to-peer network that supports an error-free broadcast channel from each user to all others. The DFL system aims to communication-efficiently aggregate the users’ inputs in the presence of dropout users, while ensuring information-theoretic privacy of each individual input against a certain number of colluding users.

Based on the absolute magnitudes of the input, a top-$K$ sparsification strategy is employed to reduce the communication overhead in the secure aggregation. More specifically, for each user $n\in[N]$, the top-$K$ sparsified representation of the input $W_n$ is characterized by an index set $\mathcal{K}_n\subseteq [L]$ with $\vert \mathcal{K}_n \vert = K$ together with the associated values $\{w_{n,k}\}_{k\in\mathcal{K}_n}$.
Here, $\mathcal{K}_n$ denotes the set of indices corresponding to the $K$ entries of $W_n$ with the largest absolute magnitudes, and $w_{n,k}$ denotes the value of $W_n$ at coordinate $k$ for any $k\in\mathcal{K}_n$.\footnote{Note that, in practice, the inputs are defined over the real numbers. Information-theoretic security requires quantizing these real-valued inputs into a finite field. In this context, the top-$K$ operation essentially refers to selecting the $K$ entries with the largest absolute magnitudes in the real domain.} 
To facilitate the aggregation, we equivalently represent the top-$K$ sparsification of $W_n$ as a sparsified vector $\mathsf{TopK}(W_n)$ of length $L$, whose $k$-th entry $\mathsf{TopK}_{n,k}$ is defined by
\begin{IEEEeqnarray}{c}\label{TopK:expression}
  \mathsf{TopK}_{n,k} = 
  \begin{cases}
    w_{n,k}, & \text{if } k \in \mathcal{K}_n \\
    0, & \text{otherwise}
  \end{cases}, \quad\forall\,k\in[L].
\end{IEEEeqnarray}
In addition, to provide privacy guarantees, each user $n\in[N]$ has a random variable $Z_n$. The randomness across all users is correlated in a specific manner to ensure the correct computation of the desired aggregation. Since these random variables are generated independently of the inputs, they can be produced offline prior to performing secure aggregation.

For robustness against user dropouts, a sparse secure aggregation scheme is organized into two phases: a \emph{masked-input phase} and a \emph{mask-elimination phase}.

\textbf{Masked-Input Phase:} Each user $n\in[N]$ locally generates a masked version $X_n$ of its sparse input $\mathsf{TopK}(W_n)$ using the available randomness $Z_n$, i.e.,
\begin{IEEEeqnarray*}{c}
  H(X_n | \mathsf{TopK}(W_n), Z_n) = 0, \quad \forall\, n \in [N].
\end{IEEEeqnarray*}
Then, user $n$ broadcasts the masked input $X_n$ to all other users. 
Some users may drop out during this phase, and thus cannot participate in the subsequent aggregation.
Let $\mathcal{U}_1 \subseteq [N]$ denote the set of surviving users, with $U \leq |\mathcal{U}_1| \leq N$, where $U$ represents the minimum number of surviving users required by the DFL system. After this communication, each surviving user $n\in\mathcal{U}_1$ has received the messages $\{X_m\}_{m\in\mathcal{U}_1\setminus \{n\}}$, with the goal of securely computing the aggregation of the sparse inputs of all surviving users, i.e., $\sum_{n\in\mathcal{U}_1}\!\!\mathsf{TopK}(W_n)$.

\textbf{Mask-Elimination Phase:} 
In order to cancel the interference introduced by the masking, each surviving user $n \!\in\! \mathcal{U}_1$ broadcasts an additional message $Y_n^{\mathcal{U}_1}$, which is a deterministic function of the currently available information, including $\mathsf{TopK}(W_n),  W_n,  Z_n$, and $\{X_m\}_{m\in\mathcal{U}_1}$, i.e.,
{
  \setlength{\abovedisplayskip}{8pt}
  \setlength{\belowdisplayskip}{8pt}
\begin{IEEEeqnarray*}{c}
  H(Y_n^{\mathcal{U}_1} \Bigm| \mathsf{TopK}(W_n),W_n, Z_n, \{X_m\}_{m\in\mathcal{U}_1}) = 0. 
\end{IEEEeqnarray*}}%
Additional user dropouts may occur during the mask elimination phase. Let $\mathcal{U}_2\!\subseteq\!\mathcal{U}_1$
denote the set of surviving users after this phase, with $U\!\leq\!\vert\mathcal{U}_2\vert\!\leq\! |\mathcal{U}_1|\!\leq\! N$.
Accordingly, each surviving user $n\!\in\!\mathcal{U}_2$ observes the messages $\{Y_m^{\mathcal{U}_1}\}_{m\in\mathcal{U}_2 \setminus \{n\}}$ and attempts to reconstruct the sparse aggregate $\sum_{n\in\mathcal{U}_1}\!\!\mathsf{TopK}(W_n)$.

The secure aggregation scheme with top-$K$ sparsification must satisfy the following information-theoretic constraints.
\begin{itemize}
    \item \textbf{Correctness:} For all possible dropout patterns with $\mathcal{U}_2\!\subseteq\!\mathcal{U}_1\!\subseteq\![N]$ satisfying $U\!\!\leq\!\!\vert\mathcal{U}_2\vert\!\!\leq\!\! |\mathcal{U}_1|\!\!\leq\!\! N$, each user $n\in\mathcal{U}_2$ can correctly decode $\sum_{n\in\mathcal{U}_1}\!\!\mathsf{TopK}(W_n)$ from all available information after the two-phase communication, i.e.,
    {
      \setlength{\abovedisplayskip}{6pt}
      \setlength{\belowdisplayskip}{6pt}
    \begin{IEEEeqnarray}{c}
    H\Big(\sum_{n\in\mathcal{U}_1}\!\mathsf{TopK}(W_n) \Big{|} \{X_n\}_{n\in\mathcal{U}_1}, \qquad\qquad\qquad\qquad\quad \nonumber  \\ 
    \qquad\quad \{Y_n^{\mathcal{U}_1}\}_{n\in\mathcal{U}_2}, \mathsf{TopK}(W_n),W_n, Z_n \Big) = 0. 
    \label{eq:correctness}
    \end{IEEEeqnarray}}%

    \item \textbf{Security:} Any group of up to $T$ colluding users must not learn any additional information about all users' inputs, beyond what is inherently revealed by the desired sparse aggregate and the colluding users’ own inputs. 
    Formally, for any subset $\mathcal{T}\subseteq[N]$ with $|\mathcal{T}|\leq T$ and any dropout pattern with $\mathcal{U}_2\!\subseteq\!\mathcal{U}_1\!\subseteq\![N]$ such that $U\!\!\leq\!\!\vert\mathcal{U}_2\vert\!\!\leq\!\! |\mathcal{U}_1|\!\!\leq\!\! N$, the following security constraint must hold:
    {
      \setlength{\abovedisplayskip}{6pt}
      \setlength{\belowdisplayskip}{6pt}
    \begin{IEEEeqnarray}{l}
    I\Big(\{W_n\}_{n\in[N]}; \{\mathsf{TopK}(W_n),Z_n\}_{n\in\mathcal{T}}, \{X_n\}_{n\in\mathcal{U}_1}, \quad \quad \nonumber \\ 
    \qquad \{Y_n^{\mathcal{U}_1}\}_{n\in\mathcal{U}_2}\Big{|}  \sum_{n\in\mathcal{U}_1} \mathsf{TopK}(W_n), \left\{W_n\right\}_{n\in\mathcal{T}}\Big) = 0. 
    \IEEEeqnarraynumspace 
    \label{eq:security}
    \end{IEEEeqnarray}}%
\end{itemize}

The communication rates characterize the number of symbols that must be transmitted per input symbol to complete the desired sparse aggregation, and are defined as follows.
{
  \setlength{\abovedisplayskip}{8pt}
  \setlength{\belowdisplayskip}{8pt}
\begin{IEEEeqnarray}{c}
R_1=\frac{\max\limits_{n\in [N]} H(X_n)}{L}, \quad
R_2=\frac{\max\limits_{n\in\mathcal{U}_1} H(Y_{n}^{\mathcal{U}_1})}{L}. \notag
\end{IEEEeqnarray}}%

The goal of this paper is to design a top-$K$ sparse secure aggregation scheme that satisfies the constraints in \eqref{eq:correctness} and \eqref{eq:security} (that is, a specification of $\{Z_n\}_{n\in[N]},\{X_n\}_{n\in[N]},\{Y_n^{\mathcal{U}_1}\}_{n\in\mathcal{U}_1}$ for all admissible $\mathcal{U}_1$), while achieving its communication rates $R_1$ and $R_2$ as small as possible.

\section{The Proposed Sparse Secure Aggregation}
\label{sec:achievability}
This section designs a top-$K$ sparse secure aggregation scheme. 
We begin with an illustrated example.

\subsection{An Illustrated Example}
Consider a DFL system with system parameters $N=5,U=3,T=1$, and $K=2$, in which each user $n\in[5]$ holds a private input $W_n$ of length $L=4$ over $\mathbb{F}_q$.
Without loss of generality, assume that the top-$2$ sparsified representations of the inputs are given by
{
  \setlength{\abovedisplayskip}{5pt}
  \setlength{\belowdisplayskip}{5pt}
\begin{IEEEeqnarray}{rl}
(\mathcal{K}_1 \!=\! \{2,4\},w_{1,2},w_{1,4}),& \quad (\mathcal{K}_2 \!=\! \{3,4\},w_{2,3},w_{2,4}), \label{example:support:1} \IEEEeqnarraynumspace \\
(\mathcal{K}_3 \!=\! \{1,3\},w_{3,1},w_{3,3}),& \quad (\mathcal{K}_4 \!=\! \{2,3\},w_{4,2},w_{4,3}),   \label{example:support:2} \\
(\mathcal{K}_5 \!=\! \{1,4\},w_{5,1},w_{5,4}). &  \label{example:support:3} 
\end{IEEEeqnarray}}%
In the DFL system with at least $U=3$ surviving users, the users aim to perform sparse secure aggregation while ensuring that the sparse gradients remain information-theoretically private against any $T=1$ colluding user.

\textbf{Offline Phase:}
This phase generates a random variable $Z_n$ for each user $n\in[5]$ to enable sparse secure aggregation.
Let $\alpha_1,\alpha_2,\alpha_3,\alpha_4,\alpha_5$ and $\beta_1,\beta_2,\beta_3$ be $N+U=8$ pairwise distinct public elements over $\mathbb{F}_q$.

To hide the support set $\mathcal{K}_n$, each user $n\in[5]$ independently selects a random permutation $\pi_n$ of $[4]$, given by
{
  \setlength{\abovedisplayskip}{5.25pt}
  \setlength{\belowdisplayskip}{5.25pt}
\begin{IEEEeqnarray*}{c}
  \pi_n = 
    \tightpmatrix{
      1 & 2 & 3 & 4 \\
      \pi_n(1) & \pi_n(2) & \pi_n(3) & \pi_n(4) }.
\end{IEEEeqnarray*}}%
Let $\sigma_n$ denote the inverse permutation of $\pi_n$ such that
{
  \setlength{\abovedisplayskip}{5pt}
  \setlength{\belowdisplayskip}{5pt}
\begin{IEEEeqnarray*}{c}
\sigma_n(\pi_n(1))=1,\quad
\sigma_n(\pi_n(2))=2,\\
\sigma_n(\pi_n(3))=3, \quad
\sigma_n(\pi_n(4))=4.
\end{IEEEeqnarray*}}%
Equivalently, $\sigma_n$ can be represented as a $4\times4$ matrix $\mathbf{P}_n$. For example, the following two forms are equivalent:
{
  \setlength{\abovedisplayskip}{6pt}
  \setlength{\belowdisplayskip}{6pt}
\begin{IEEEeqnarray*}{c}
\sigma_n = 
    \tightpmatrix{
      1 & 2 & 3 & 4 \\
      3 & 4 & 1 & 2},\quad
  \mathbf{P}_n = 
    \tightbmatrix{
      0 & 0 & 1 & 0 \\
      0 & 0 & 0 & 1 \\
      1 & 0 & 0 & 0 \\
      0 & 1 & 0 & 0}.
\end{IEEEeqnarray*}}%
Let $\mathbf{p}_{n,i}$ denote the $i$-th row vector of the matrix $\mathbf{P}_n$ for any $i\in[4]$ and $n\in[5]$. Obviously, we have $\mathbf{p}_{n,\pi_n(i)} = \mathbf{e}_i$ for any $i\in[4]$, where $\mathbf{e}_i$ denotes a length-$4$ vector whose $i$-th entry is $1$ and all other entries are $0$.

The user $n\in[5]$ independently and uniformly samples a random vector $\mathbf{r}_n\!=\!(r_{n,1}, r_{n,2}, r_{n,3}, r_{n,4})$ from $\mathbb{F}_q^4$ and applies the local permutation $\sigma_n$ to obtain the permuted vector $(r_{n,\sigma_n(1)},\! r_{n,\sigma_n(2)},\, r_{n,\sigma_n(3)},\, r_{n,\sigma_n(4)})$.
Then, the vectors $\mathbf{p}_{n,i}$ and $r_{n,\sigma_n(i)}\mathbf{p}_{n,i}$ are secretly shared with the other users. 
More precisely, to completely exploit the $U=3$ surviving users to reduce the communication cost in the mask-elimination phase, $\mathbf{p}_{n,i}$ is partitioned into $D=U-T=2$ blocks, given by
\begin{IEEEeqnarray}{c}\label{example:partition}
\mathbf{p}_{n,i}=\big(\mathbf{p}_{n,i}^{(1)},\mathbf{p}_{n,i}^{(2)}\big),\quad
\forall\, i\in[4],n\in[5].
\end{IEEEeqnarray}
For each $i\in[4]$, the user $n$ constructs two secure Lagrange interpolation polynomials $f_{n,i}(x)$ and $h_{n,i}(x)$, each of degree $D+T-1=2$, satisfying
{
  \setlength{\abovedisplayskip}{5pt}
  \setlength{\belowdisplayskip}{5pt}
\begin{IEEEeqnarray*}{l}
f_{n,i}(\beta_1)=\mathbf{p}_{n,i}^{(1)}, \quad f_{n,i}(\beta_2)=\mathbf{p}_{n,i}^{(2)}, \quad f_{n,i}(\beta_3)=\mathbf{z}_{n,i}, \IEEEeqnarraynumspace \\
h_{n,i}(\beta_1)=r_{n,\sigma_n(i)}\mathbf{p}_{n,i}^{(1)}, \quad  h_{n,i}(\beta_2)=r_{n,\sigma_n(i)}\mathbf{p}_{n,i}^{(2)}, \\
h_{n,i}(\beta_3)=\tilde{\mathbf{z}}_{n,i},
\end{IEEEeqnarray*}}%
where $\mathbf{z}_{n,i}$ and $\tilde{\mathbf{z}}_{n,i}$ are the random noise vectors used to provide $T=1$-privacy guarantees. 

Next, the user $n\in[5]$ evaluates the polynomials $f_{n,i}(x)$ and $h_{n,i}(x)$ at $x=\alpha_m$ and then sends the resulting evaluations to user $m\in[5]$ for all $i\in[4]$.
Collecting all the information generated and received in this offline phase, the user $n$ stores the random variable
{
  \setlength{\abovedisplayskip}{4pt}
  \setlength{\belowdisplayskip}{4pt}
\begin{IEEEeqnarray*}{c}
{Z}_n\!\!=\!\!\big\{\!\pi_n,\! \{f_{m,i}(\alpha_n),h_{m,i}(\alpha_n)\}_{m\in[5],i\in[4]},\!\mathbf{r}_n,\! \{\mathbf{z}_{n,i},\tilde{\mathbf{z}}_{n,i}\}_{i\in[4]}\!\big\}.\!
\end{IEEEeqnarray*}}%
This random variable $Z_n$ captures the offline view of user~$n$ and is independent of the private input $W_n$.

\textbf{Masked-Input Phase:} Each user $n$ masks its private support set $\mathcal{K}_n$ using the random permutation $\pi_n$ and masks the corresponding gradient values $\{w_{n,k}\}_{k\in\mathcal{K}_n}$ using the random noises $\{r_{n,k}\}_{k\in\mathcal{K}_n}$. For all $n\in[5]$, by \eqref{example:support:1}--\eqref{example:support:3}, the masked message $X_n$ broadcast by user $n$ to all users is designed as
{
  \setlength{\abovedisplayskip}{4pt}
  \setlength{\belowdisplayskip}{4pt}
\begin{IEEEeqnarray*}{rCl}
{X}_1 = \{\pi_1(2),\pi_1(4), w_{1,2}+r_{1,2},w_{1,4}+r_{1,4}\}, 
  \label{eq:design_X1} \\
{X}_2 = \{\pi_2(3),\pi_2(4), w_{2,3}+r_{2,3},w_{2,4}+r_{2,4}\}, 
  \label{eq:design_X2} \\
{X}_3 = \{\pi_3(1),\pi_3(3), w_{3,1}+r_{3,1},w_{3,3}+r_{3,3}\}, 
  \label{eq:design_X3} \\
{X}_4 = \{\pi_4(2),\pi_4(3), w_{4,2}+r_{4,2},w_{4,3}+r_{4,3}\},
  \label{eq:design_X4} \\
{X}_5 = \{\pi_5(1),\pi_5(4), w_{5,1}+r_{5,1},w_{5,4}+r_{5,4}\}.
  \label{eq:design_X5}
\end{IEEEeqnarray*}}%

Without loss of generality, suppose that user $5$ drops out after this phase, i.e., the surviving user set is $\mathcal{U}_1=\{1,2,3,4\}$. Each surviving user wishes to complete the sparse aggregation $(w_{3,1},w_{1,2}+w_{4,2},w_{2,3}+w_{3,3}+w_{4,3},w_{1,4}+w_{2,4})$ by \eqref{example:support:1}-\eqref{example:support:2}.

\textbf{Mask-Elimination Phase:} To eliminate the interference introduced by the random masks, upon receiving the masked messages $\{X_m\}_{m\in\mathcal{U}_1}$, the surviving user $n\in\{1,2,3,4\}$ generates an additional message $Y_n^{\mathcal{U}_1}$ using its local randomness $Z_n$, given by
{
  \setlength{\abovedisplayskip}{4pt}
  \setlength{\belowdisplayskip}{3pt}
\begin{IEEEeqnarray*}{c}
Y_n^{\mathcal{U}_1}\!\!=\!\!\!\sum_{m\in\mathcal{U}_1}\! \sum_{k\in\mathcal{K}_m}
     \!\!\!\! \Big(\!(w_{m,k}\!+\!r_{m,k}) f_{m,\pi_m(k)}(\alpha_n) \!-\! h_{m,\pi_m(k)}(\alpha_n)\!\Big),
\end{IEEEeqnarray*}}%
and broadcasts it to the other surviving users.

Following the mask-elimination phase, suppose further that user $4$ drops out, so that the surviving user set in this phase is $\mathcal{U}_2 =\{1,2,3\}$. Therefore, each surviving user $n\in\{1,2,3\}$ receives the messages $\{Y_n^{\mathcal{U}_1}\}_{n\in\{1,2,3\}}$. Let $Y^{\mathcal{U}_1}(x)$ denote a polynomial of degree at most $2$, given by
{
  \setlength{\abovedisplayskip}{4pt}
  \setlength{\belowdisplayskip}{3pt}
\begin{IEEEeqnarray*}{c}
Y^{\mathcal{U}_1}\!(x)\!\!=\!\!\!\sum_{m\in\mathcal{U}_1}\! \sum_{k\in\mathcal{K}_m}
     \!\!\!\! \Big(\!(w_{m,k}\!+\!r_{m,k}) f_{m,\pi_m(k)}(x) \!-\! h_{m,\pi_m(k)}(x)\!\Big).
\end{IEEEeqnarray*}}%
Since the received message $Y_n^{\mathcal{U}_1}$ is an evaluation of the polynomial $Y^{\mathcal{U}_1}(x)$ at $x=\alpha_n$ for $n\in\{1,2,3\}$, each surviving user can interpolate $Y^{\mathcal{U}_1}(x)$ from the messages $\{Y_n^{\mathcal{U}_1}\}_{n\in\{1,2,3\}}$. Evaluating the interpolated polynomial at $\beta_1$ and $\beta_2$ yields
{
  \setlength{\abovedisplayskip}{4pt}
  \setlength{\belowdisplayskip}{3pt}
\begin{IEEEeqnarray*}{rCl}
Y^{\mathcal{U}_1}(\beta_1)&=&\sum_{m\in\mathcal{U}_1}\sum_{k\in\mathcal{K}_m} w_{m,k}\mathbf{p}_{m,\pi_m(k)}^{(1)}, \notag\\
Y^{\mathcal{U}_1}(\beta_2)&=&\sum_{m\in\mathcal{U}_1}\sum_{k\in\mathcal{K}_m} w_{m,k}\mathbf{p}_{m,\pi_m(k)}^{(2)}.
\end{IEEEeqnarray*}}%
By concatenating the two decoding data $Y^{\mathcal{U}_1}(\beta_1)$ and $Y^{\mathcal{U}_1}(\beta_2)$ according to \eqref{example:partition}, the user computes the sparse aggregate:
{
  \setlength{\abovedisplayskip}{4pt}
  \setlength{\belowdisplayskip}{4pt}
\begin{IEEEeqnarray*}{l}
\sum_{m\in\mathcal{U}_1}\sum_{k\in\mathcal{K}_m} w_{m,k}\mathbf{p}_{m,\pi_m(k)}=
\sum_{m\in\mathcal{U}_1}\sum_{k\in\mathcal{K}_m} w_{m,k}\mathbf{e}_k= \notag \\
\quad\quad (w_{3,1},w_{1,2}+w_{4,2},w_{2,3}+w_{3,3}+w_{4,3},w_{1,4}+w_{2,4}). \notag
\end{IEEEeqnarray*}}%

In this example,
each masked-input message consists of $K=2$ masked symbols along with $K=2$ distinct indices chosen from $L=4$ indices, resulting in a communication rate of $R_1 = (2 + \log_q \binom{4}{2})/4$.
In the mask-elimination phase, each survivor broadcasts a length-$2$
vector, yielding a communication rate of $R_2=1/2$.

\subsection{General Construction of Sparse Secure Aggregation}
We present the general construction for any system parameters $N, U, T$, and $K$ satisfying $1 \le T < U \le N$.

\textbf{Offline Phase:} In this phase, each user $n\in[N]$ communicates with other users to initialize a random variable $Z_n$, which is used to enable sparse secure aggregation while providing information-theoretic privacy guarantees.
Let $\{\alpha_n\}_{n\in[N]}$ and $\{\beta_{u}\}_{u\in[U]}$ be $N+U$ pairwise distinct elements from $\mathbb{F}_q$, which serve as the encoding parameters 
and are publicly known to all users. This also means that the finite field satisfies $q\geq N+U$.

To protect the privacy of the top-$K$ indices, each user $n\in[N]$ independently samples a uniformly random permutation $\pi_n$ on the index set $[L]$, given by
{
  \setlength{\abovedisplayskip}{4pt}
  \setlength{\belowdisplayskip}{4pt}
\begin{IEEEeqnarray*}{c}
\pi_n= 
\begin{pmatrix}
1 & 2 & \cdots & L \\
\pi_n(1) & \pi_n(2) & \cdots & \pi_n(L)
\end{pmatrix},\quad\forall\,n\in[N]. \IEEEeqnarraynumspace
\end{IEEEeqnarray*}}%
Let $\sigma_n$ denote the inverse permutation of $\pi_n$, i.e.,
{
  \setlength{\abovedisplayskip}{4pt}
  \setlength{\belowdisplayskip}{4pt}
\begin{IEEEeqnarray}{c}\label{inverse:permutation}
\sigma_n(\pi_n(k))=k,\quad\forall\, k\in[L].
\end{IEEEeqnarray}}%
This inverse permutation $\sigma_n$ can be equivalently represented by an $L\times L$ matrix $\mathbf{P}_{n}$, whose entries are defined as
{
  \setlength{\abovedisplayskip}{4pt}
  \setlength{\belowdisplayskip}{4pt}
\begin{IEEEeqnarray}{c}\label{permutation:matrix}
  p_{n,i,j} = 
  \begin{cases}
    1, & \text{if } j =\sigma_n(i) \\
    0, & \text{otherwise}
  \end{cases}, \quad\forall\,i,j\in[L],
\end{IEEEeqnarray}}%
where $p_{n,i,j}$ denotes the entry in the $i$-th row and $j$-th column of the matrix $\mathbf{P}_{n}$.
Then, the user $n\in[N]$ secretly shares the random permutation matrix $\mathbf{P}_{n}$ with the other users. 
To fully leverage the $U$ surviving users to reduce the communication overhead in the following mask-elimination phase, we additionally introduce the hyperparameter $D$, which must satisfy the constraint $D+T\leq U$ to ensure decodability while providing $T$-colluding privacy guarantees.

Specifically, let $\mathbf{p}_{n,i}$ denote the $i$-th row vector of matrix $\mathbf{P}_{n}$ for $i\in[L]$ and $n\in[N]$. 
The vector $\mathbf{p}_{n,i}$ is partitioned into $D$ equal-sized subvectors of each length $L/D$,\footnote{Here, it is implicitly assumed that $D$ divides $L$.} given by
{
  \setlength{\abovedisplayskip}{5pt}
  \setlength{\belowdisplayskip}{5pt}
\begin{IEEEeqnarray}{c}\label{partition}
\mathbf{p}_{n,i}=\big(\mathbf{p}_{n,i}^{(1)},\mathbf{p}_{n,i}^{(2)},\ldots,\mathbf{p}_{n,i}^{(D)}\big),\quad\forall\, i\in[L],n\in[N].\IEEEeqnarraynumspace
\end{IEEEeqnarray}}%
The user $n\in[N]$ locally selects a random vector $\mathbf{r}_n$ of length $L$ over $\mathbb{F}_q$ and applies the inverse permutation $\sigma_n$ to $\mathbf{r}_n$. 
Let $r_{n,i}$ denote the $i$-th entry of $\mathbf{r}_n$ for any $i\in[L]$, so that $r_{n,\sigma_n(i)}$ is the $i$-th entry of the permuted vector.
Then, for each $i\in[L]$, the user $n$ secretly shares the two vectors $\mathbf{p}_{n,i}$ and $r_{n,\sigma_n(i)}\mathbf{p}_{n,i}$ with the other users. 
This is achieved by constructing two secure Lagrange encoding polynomials $f_{n,i}(x)$ and $h_{n,i}(x)$ \cite{yu2019lagrange,zhu2022symmetric,zhu2022generalized}, each of degree at most $D+T-1$, satisfying
{
  \setlength{\abovedisplayskip}{4pt}
  \setlength{\belowdisplayskip}{4pt}
\begin{IEEEeqnarray}{rCl}
  f_{n,i}(\beta_d) &=& 
  \begin{cases}
    \mathbf{p}_{n,i}^{(d)}, & \text{if } d\in[D] \\
    \mathbf{z}_{n,i}^{(d)}, & \text{if } d\in[D+1:D+T]
  \end{cases}, \label{polynomial:1} \\
  h_{n,i}(\beta_d) &=& 
  \begin{cases}
    r_{n,\sigma_n(i)} \mathbf{p}_{n,i}^{(d)}, & \text{if } d\in[D] \\
    \tilde{\mathbf{z}}_{n,i}^{(d)}, & \text{if } d\in[D+1:D+T] 
  \end{cases}, \IEEEeqnarraynumspace \label{polynomial:2}
\end{IEEEeqnarray}}%
where $\{\mathbf{z}_{n,i}^{(d)},\tilde{\mathbf{z}}_{n,i}^{(d)}\}_{d\in[D+1:D+T]}$ are random noises from $\mathbb{F}_q^{L/D}$ used to provide privacy guarantees.

For any $n\in[N]$ and $i\in[L]$, the polynomials $f_{n,i}(x)$ and $h_{n,i}(x)$ are evaluated at $x=\alpha_m$ and the resulting values are securely sent to user $m$ for all $m\in[N]$. 
Consequently, for any $n\in[N]$, the random variable ${Z}_n$ accessible to the user $n$ is given by
{
  \setlength{\abovedisplayskip}{5pt}
  \setlength{\belowdisplayskip}{5pt}
\begin{IEEEeqnarray}{l}
{Z}_n=\big\{\pi_n,\{f_{m,i}(\alpha_n),h_{m,i}(\alpha_n)\}_{m\in[N],i\in[L]},\notag\\
\qquad\qquad\qquad\qquad \mathbf{r}_n,\{\mathbf{z}_{n,i}^{(d)},\tilde{\mathbf{z}}_{n,i}^{(d)}\}_{d\in[D+1:D+T],i\in[L]}\big\}. \IEEEeqnarraynumspace \label{local:noise}
\end{IEEEeqnarray}}%

\textbf{Masked-Input Phase:}
Given its private input $W_n$, user $n\in[N]$ can determine the support set $\mathcal{K}_n \subseteq [L]$ with $\vert\mathcal{K}_n\vert = K$ corresponding to the $K$ entries with the largest absolute magnitudes, along with the associated values $\{w_{n,k}\}_{k\in\mathcal{K}_n}$.

To provide privacy guarantees, the local available random permutation $\pi_n$ is employed to obfuscate the indices in $\mathcal{K}_n$, and the local available random noises $\{r_{n,k}\}_{k\in\mathcal{K}_n}$ are used to mask the sparse values $\{w_{n,k}\}_{k\in\mathcal{K}_n}$, given by
{
  \setlength{\abovedisplayskip}{4pt}
  \setlength{\belowdisplayskip}{3pt}
\begin{IEEEeqnarray}{c}\label{message:first}
x_{n,k}=w_{n,k}+r_{n,k},\quad\forall\,k\in\mathcal{K}_n.
\end{IEEEeqnarray}}%
Therefore, the user $n$ generates the masked message ${X}_n$ as
{
  \setlength{\abovedisplayskip}{4pt}
  \setlength{\belowdisplayskip}{3pt}
\begin{IEEEeqnarray}{c}\label{message:phase:1}
{X}_n = \{(\pi_n(k), x_{n,k}) : k\in\mathcal{K}_n \},\quad\forall\,n\in[N],
\end{IEEEeqnarray}}%
and then broadcasts it to all the other users. 

After this communication, each surviving user can identify the set $\mathcal{U}_1$ of all surviving users and obtain the masked messages $\{{X}_n\}_{n\in\mathcal{U}_1}$, with the goal of recovering the desired sparse aggregation $\sum_{n\in\mathcal{U}_1}\!\!\mathsf{TopK}(W_n)$.

\setlength{\parskip}{1.2pt}

\textbf{Mask-Elimination Phase:} Based on the available randomness ${Z}_n$ in \eqref{local:noise} and the received messages $\{{X}_n\}_{n\in\mathcal{U}_1}$ in \eqref{message:phase:1}, each surviving user $n\in\mathcal{U}_1$ can generate $Y_n^{\mathcal{U}_1}$ as
{
  \setlength{\abovedisplayskip}{5pt}
  \setlength{\belowdisplayskip}{4pt}
\begin{IEEEeqnarray}{c}\label{second:message}
Y_n^{\mathcal{U}_1}\!\!=\!\!\sum\limits_{m\in\mathcal{U}_1}\sum\limits_{k\in\mathcal{K}_m}\!\!\!\Big(x_{m,k}\!\cdot\! f_{m,\pi_m(k)}(\alpha_n)-h_{m,\pi_m(k)}(\alpha_n)\Big), \IEEEeqnarraynumspace
\end{IEEEeqnarray}}%
and then broadcasts $Y_n^{\mathcal{U}_1}$ to all other users. Accordingly, in the communication phase, the surviving user $n\in\mathcal{U}_2$ will receive the messages $\{Y_m^{\mathcal{U}_1}\}_{m\in\mathcal{U}_2}$ for any $\mathcal{U}_2\subseteq\mathcal{U}_1\subseteq[N]$.

Next, we prove that the proposed sparse secure aggregation scheme satisfies the correctness constraint in \eqref{eq:correctness}. It suffices to show that each surviving user $n\in\mathcal{U}_2$ can recover the desired sparse aggregation $\sum_{n\in\mathcal{U}_1}\!\!\mathsf{TopK}(W_n)$ from the received messages $\{Y_m^{\mathcal{U}_1}\}_{m\in\mathcal{U}_2}$ for any surviving sets subject to $U\leq\vert\mathcal{U}_2\vert\leq |\mathcal{U}_1|\leq N$ and $D+T\leq U$.

Let $Y^{\mathcal{U}_1}(x)$ be a degree $D+T-1$ polynomial, given by
{
  \setlength{\abovedisplayskip}{4pt}
  \setlength{\belowdisplayskip}{3pt}
\begin{IEEEeqnarray*}{rCl}
Y^{\mathcal{U}_1}(x)\! &=& \!\sum\limits_{m\in\mathcal{U}_1}\sum\limits_{k\in\mathcal{K}_m}\!\!\!\Big(x_{m,k}\!\cdot\! f_{m,\pi_m(k)}(x)-h_{m,\pi_m(k)}(x)\Big). 
\end{IEEEeqnarray*}}%
Observe that, the received message $Y_n^{\mathcal{U}_1}$ can be viewed as an evaluation of $Y^{\mathcal{U}_1}(x)$ at $x=\alpha_n$ for any $n\in\mathcal{U}_2$. Due to the constraint $D+T\leq U$, the user $n$ can recover $Y^{\mathcal{U}_1}(x)$ from the received messages $\{Y_m^{\mathcal{U}_1}\}_{m\in\mathcal{U}_2}$ via Lagrange polynomial interpolation. Then, by \eqref{inverse:permutation}, \eqref{polynomial:1}, \eqref{polynomial:2}, and \eqref{message:first}, evaluating $Y^{\mathcal{U}_1}(x)$ at $x=\beta_d$ for all $d\in[D]$ yields
{
  \setlength{\abovedisplayskip}{4pt}
  \setlength{\belowdisplayskip}{3pt}
\begin{IEEEeqnarray}{rCl}
Y^{\mathcal{U}_1}(\beta_d)
&=&\sum\limits_{m\in\mathcal{U}_1}\sum\limits_{k\in\mathcal{K}_m}\big(w_{m,k}\cdot \mathbf{p}_{m,\pi_m(k)}^{(d)} \big). \notag 
\end{IEEEeqnarray}}%
By stacking these terms $\sum_{m\in\mathcal{U}_1}\sum_{k\in\mathcal{K}_m}(w_{m,k}\cdot \mathbf{p}_{m,\pi_m(k)}^{(d)})$ over all $d\in[D]$ in accordance with \eqref{partition},
each surviving user $n\in\mathcal{U}_2$ can recover the desired sparse aggregation $\sum_{n\in\mathcal{U}_1}\!\!\mathsf{TopK}(W_n)$ as follows:
{
  \setlength{\abovedisplayskip}{4pt}
  \setlength{\belowdisplayskip}{3pt}
\begin{IEEEeqnarray}{rCl}
\sum\limits_{m\in\mathcal{U}_1}\sum\limits_{k\in\mathcal{K}_m}(w_{m,k}\cdot \mathbf{p}_{m,\pi_m(k)} )&=& \sum\limits_{m\in\mathcal{U}_1}\sum\limits_{k\in\mathcal{K}_m}(w_{m,k}\cdot \mathbf{e}_k ) \notag \\
&=&\sum\limits_{m\in\mathcal{U}_1}\!\!\mathsf{TopK}(W_m),
\notag \IEEEeqnarraynumspace \notag 
\end{IEEEeqnarray}}%
where the first equation follows from \eqref{inverse:permutation}-\eqref{permutation:matrix},
and the second equation is due to \eqref{TopK:expression}.

\emph{Performance Analysis:}
In the masked-input phase in \eqref{message:phase:1}, each user broadcasts $K$ masked symbols together with an obfuscated support set of size $K$ selected from $\binom{L}{K}$ possibilities, leading to a communication rate of $R_1 = (K + \log_q \binom{L}{K})/L$.
In the mask-elimination phase in \eqref{second:message}, each surviving user broadcasts a single vector $Y_n^{\mathcal{U}_1}$ of length $\frac{L}{D}$, resulting in a rate of $R_2 = \frac{1}{U-T}$, where $D$ is set to its maximum value $U-T$.

\emph{Security Analysis:}
The proposed sparse scheme consists of three communication phases: (1) In the offline phase, the data $\{f_{n,i}(\alpha_m),h_{n,i}(\alpha_m)\}$ shared with user $m$ are protected using $T$ additional independent random noises for any $n,m\in[N]$ and $i\in[L]$. This ensures that each user’s random permutation $\pi_n$ and private mask $\mathbf{r}_n$ remain information-theoretically secure from any $T$ colluding users. (2) In the masked-input phase, the support set $\mathcal{K}_n$ of the top-$K$ entries is concealed via the random permutation $\pi_n$, while the corresponding sparse gradient values $\{w_{n,k}\}_{k\in\mathcal{K}_n}$ are masked using the random variable $\mathbf{r}_n$. Both mechanisms are analogous to a one-time pad. 
(3) In the mask-elimination phase, the messages $\{Y_{n}^{\mathcal{U}_1}\}_{n\in\mathcal{U}_2}$ observed by any $T$ colluding users are essentially functions of the data revealed in the previous two phases along with the aggregation result $\sum_{n\in\mathcal{U}_1}\!\!\!\mathsf{TopK}(W_n)$.
The above ensures that all communicated messages leak no information to any $T$ colluding users beyond the aggregation result.
Moreover, the collection of $\{\mathsf{TopK}(W_n), Z_n\}$ associated with any $T$ colluding users is independent of $\{W_n\}_{n\in[N]}$ conditioned on their own inputs, the aggregation result, and all other information obtained throughout the communication of the scheme.
Taken together, the security definition in \eqref{eq:security} is satisfied.

\vspace{-0.2em}
\begin{Theorem}
 \label{thm:achievability}
For the information-theoretic secure aggregation problem with top-$K$ sparsification considered in this paper, there exists a sparse secure aggregation scheme achieving the following communication rates for any system parameters $N,U,T,K,L$, and $q$ satisfying $1 \!\leq\! T\! <\! U\! \leq\! N$ and $q\geq N+U$:
{
  \setlength{\abovedisplayskip}{4pt}
  \setlength{\belowdisplayskip}{3pt}
  \begin{IEEEeqnarray}{c}
    R_1 = \frac{K + \log_q \binom{L}{K}}{L}, \quad R_2 = \frac{1}{U-T}. \notag
  \end{IEEEeqnarray}}%
\end{Theorem}

\vspace{-0.5em}
\begin{Remark}[Connection with Centralized Federated Learning]
Although we only focus on the sparse secure aggregation problem in decentralized federated learning, it is clear that the proposed sparse scheme can be directly applied to the conventional centralized federated learning system with a single central server \cite{bonawitz2017practical,so2022lightsecagg, zhao2022information}. In the centralized setting, the sparse aggregation result can be decoded at the server, which then broadcasts the decoded result to the users, while all other steps remain unchanged. The security guarantee also naturally holds, since the joint view of the server and any $T$ colluding users in the centralized system is identical to the information observed by any $T$ users in the decentralized setting.
\end{Remark}

\vspace{-0.5em}
\begin{Remark}[Efficiency Optimization]
While we propose the first top-$K$ sparse information-theoretic secure aggregation scheme,
it requires secretly sharing full-dimensional permutation matrices, which incurs an offline overhead that scales with the gradient dimension $L$.
To enhance scalability for large-scale models, one potential direction for future research is to relax information-theoretic security to computational security by using shared short cryptographic seeds \cite{bonawitz2017practical, bell2020secure, Kadhe:ICML2020}.
\end{Remark}

\vspace{-0.4em}
\subsection{Experiments}
\begin{figure}[htbp]
\centering 
\vspace{-0.8em}
\includegraphics[width=0.90\linewidth]{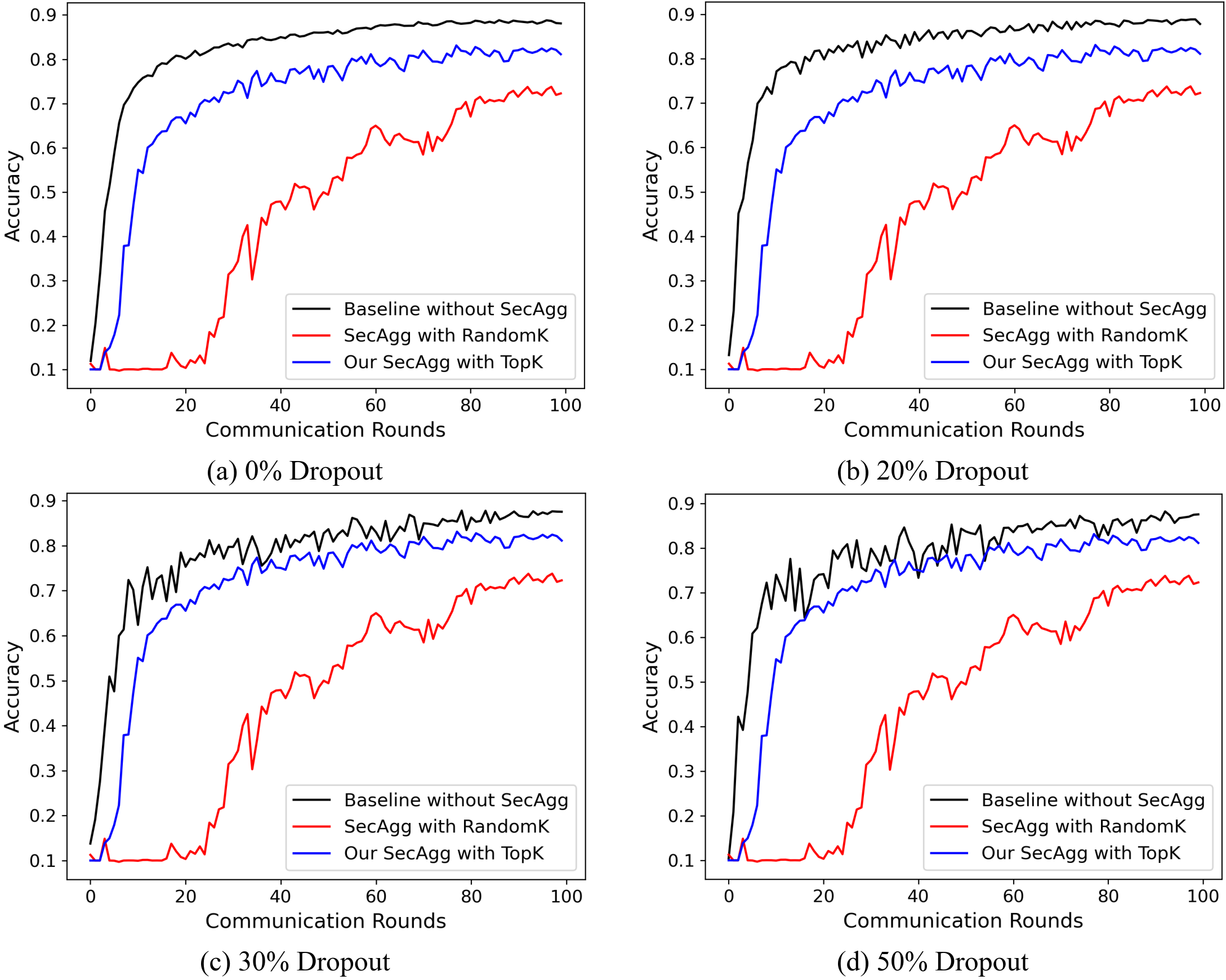}
\vspace{-0.8em}
\caption{Test accuracy under varying user dropout rates.}
\label{fig:accuracy_comparison}
\end{figure}

\vspace{-0.2em}
We evaluate the proposed top-$K$ sparse secure aggregation (SecAgg) scheme and compare it with two traditional methods: baseline without SecAgg \cite{lalitha2018fully, mcmahan2017communication} and SecAgg with random-$K$ sparsification \cite{sami2024secure}. 
Experiments are conducted on the CIFAR-10 dataset using a ResNet-20 model with $N=10$ users under a non-IID data distribution. The system parameters are set to $U\!=\!5,T\!=\!3$, and $K\!=\!0.01L$.
As shown in Fig. \ref{fig:accuracy_comparison}, under dropout rates ranging from 0\% to 50\%, our scheme achieves performance closely matching that of the baseline without SecAgg, while substantially outperforming SecAgg with Random-$K$ in terms of both convergence and accuracy, thereby demonstrating strong robustness to user dropouts.

\vspace{-0.85em}
\section{Conclusion}
\vspace{-0.1em}
In this paper, we studied information-theoretic secure aggregation with top-$K$ sparsification in decentralized federated learning under user dropouts and user collusion.
To address the challenges posed by dimension-dependent communication costs and user-specific sparsity patterns, we proposed a top-$K$ sparse secure aggregation scheme that offloads dense communication to an offline phase and employs random masks and permutations to protect private gradients.

\vspace{-0.2em}
\section*{Acknowledgment}
\vspace{-0.1em}
This work was supported in part by the National Natural Science Foundation of China (NSFC) under Grant 62401482 and Grant 62371401 and in part by the Sichuan Science and Technology Program under Grant 2026NSFSC1428.

\bibliographystyle{IEEEtran}
\bibliography{references}
\end{document}